\documentclass[showpacs,aps,prd,reprint,superscriptaddress,nofootinbib,longbibliography]{revtex4-2}
\usepackage[colorlinks=true, pdfstartview=FitV, linkcolor=magenta,citecolor=blue, urlcolor=magenta,
bookmarks=true, bookmarksnumbered=true, breaklinks]{hyperref}
\usepackage[dvipdfmx]{graphicx}
\usepackage{amsmath,amssymb,bm,color,longtable,mathrsfs,amsfonts,slashed,ulem}

\newcommand \beq{\begin{eqnarray}}
\newcommand \eeq{\end{eqnarray}}

\newcommand{\Slash}[1]{{\ooalign{\hfil/\hfil\crcr$#1$}}}

\newcommand{\Nc}{N_{\rm c}}
\newcommand{\Nf}{N_{\rm f}}

\newcommand{\lqcd}{\Lambda_{\rm QCD}}
\newcommand{\vp}{ {\bm p}}
\newcommand{\vq}{ {\bm q}}

\newcommand{\vk}{ {\bm k}}
\newcommand{\vK}{{\bm K}}
\newcommand{\vl}{ {\bm l}}
\newcommand{\la}{\langle}
\newcommand{\ra}{\rangle}

\newcommand{\calL}{\mathcal{L}}

\newcommand{\calA}{\mathcal{A}}
\newcommand{\calB}{\mathcal{B}}

\newcommand{\calN}{\mathcal{N}}

\newcommand{\rmd}{\mathrm{d}}
\newcommand{\rmi}{\mathrm{i}}
\newcommand{\rme}{\mathrm{e}}

\newcommand{\up}{\uparrow}
\newcommand{\down}{\downarrow}

\begin{document}
\begin{flushright}
\end{flushright}

\title{Peaks of sound velocity in two color dense QCD: \\
quark saturation effects and semishort range correlations}

\author{Toru Kojo}
\email{torujj@mail.ccnu.edu.cn}
\affiliation{Key Laboratory of Quark and Lepton Physics (MOE) and Institute of Particle Physics, Central China Normal University, Wuhan 430079, China}

\author{Daiki Suenaga}
\email{suenaga@rcnp.osaka-u.ac.jp}
\affiliation{Research Center for Nuclear Physics, Osaka University, Ibaraki, 567-0048, Japan}

\date{\today}

\begin{abstract}
We discuss stiffening of  dense matter in two color QCD (QC$_2$D) where hadrons are mesons and diquark baryons.
We study two models which describe a transition of matter from the Bose-Einstein-Condensation regime at low density to the Bardeen-Cooper-Schrieffer regime at high density.
The first model is based on coherent states of diquarks, and the second is the Nambu-Jona-Lasinio model with diquark pairing terms.
We particularly focus on how quark states are occupied as baryon density increases.
We find that, due to the occupied quark levels, the ideal gas picture of diquarks breaks down at density significantly less than the density where baryon cores overlap.
The saturation of quark states at low momenta stiffens equations of state.
We also study the effects of interactions which depend on the quark occupation probability.
We argue that equations of state become very stiff when the bulk part of the quark Fermi sea has the effective repulsion but the Fermi surface enjoys the attractive correlations.
This disparity for different momentum domains is possible due to the strong channel dependence in gluon exchanges with momentum transfer of $0.2-1$ GeV.
These concepts can be transferred from QC$_2$D to QCD in any numbers of colors.

\end{abstract}


\maketitle

\section{Introduction}
\label{sec:Introduction}

The quantum chromodynamics (QCD) with two colors $\Nc=2$ (QC$_2$D) has been a useful laboratory to test theoretical conceptions 
in dense matter \cite{Kogut:1999iv,Kogut:2000ek,Splittorff:2000mm,Kanazawa:2009ks,Sun:2007fc,He:2010nb,Ratti:2004ra,Brauner:2009gu,Andersen:2010vu,Andersen:2015sma,Imai:2012hr,Suenaga:2021bjz,Kojo:2021knn,Suenaga:2019jjv,Kojo:2014vja,Contant:2019lwf,Contant:2017gtz,Strodthoff:2013cua,Strodthoff:2011tz}.
In this theory, hadrons are mesons and diquark baryons. 
At low baryon density, the dilute matter is dominated by diquark baryons, and it continuously transforms into a matter dominated by quarks.
In the literatures, this continuous transformation is often referred as the crossover from Bose-Einstein-Condensation (BEC) phase to Bardeen-Cooper-Schrieffer (BCS) phase \cite{Leggett,schrieffer1999theory,BCS-BEC_Parish}.
For even flavors, the path-integral of QC$_2$D has the positive measure so that the lattice Monte Carlo simulations are 
doable \cite{Boz:2019enj,Boz:2013rca,Cotter:2012mb,Hands:2011ye,Hands:2010gd,Hands:2007uc,Hands:2006ve,Iida:2019rah,Iida:2020emi,Muroya:2002ry,Makiyama:2015uwa,Astrakhantsev:2020tdl,Astrakhantsev:2018uzd,Bornyakov:2020kyz,Bornyakov:2017txe,Braguta:2016cpw,Braguta:2015zta,Braguta:2014gea,Boz:2018crd,Buividovich:2020gnl}.
In this paper, we use QC$_2$D in two flavors ($\Nf=2$) to discuss some concepts which have been proposed for QCD to account for the neutron star (NS) phenomenology.

The quark-hadron continuity realized in QC$_2$D can give important clues to understand recent findings in 
NS observations \cite{Arzoumanian:2017puf,Fonseca:2016tux,Demorest:2010bx,Cromartie:2019kug,Fonseca:2021wxt,Antoniadis:2013pzd,TheLIGOScientific:2017qsa,Miller:2019cac,Riley:2019yda,Miller:2021qha,Riley:2021pdl,Raaijmakers:2021uju}.
Recent analyses of NSs, with similar radii ($\simeq 12.4$ km) for $2.1$ and $1.4$ solar mass NSs \cite{Miller:2021qha,Riley:2021pdl,Raaijmakers:2021uju}, 
imply that strong first order phase transitions are disfavored for a density between nuclear saturation density $n_0$ ($\sim 0.16\, {\rm fm}^{-3}$) and the core baryon density realized in two-solar mass NSs, $n_B = 4-7n_0$.
For this reason the quark-hadron continuity (modulo weak first order transitions)  
can be a useful baseline to describe the QCD equation of state (EoS) \cite{Kojo:2020krb},
and it is important to understand how such a continuous transition proceeds microscopically.

In QCD, early constructions on NS EoS with the continuity scenario were largely based on phenomenological interpolations between hadronic and 
quark EoS \cite{Masuda:2012ed,Masuda:2015kha,Masuda:2015wva,Masuda:2012kf,Kojo:2014rca,Kojo:2015fua,Fukushima:2015bda,Baym:2017whm,Baym:2019iky,Kojo:2021wax,Ayriyan:2021prr,Minamikawa:2020jfj,Minamikawa:2021fln}. 
They found several qualitative trends, such as a peak in the sound velocity, relatively large strangeness fraction at $n_B \simeq 5n_0$, 
and the importance of semi-short range correlations. 
While these trends in QCD remain conjectures, in QC$_2$D one can directly test these concepts on the lattice.
In this paper we use QC$_2$D to discuss how stiffening of matter takes place in the quark-hadron continuity.

Rapid stiffening of dense matter was discussed by McLerran and Reddy in a quarkyonic matter model \cite{McLerran:2018hbz}.
Quarkyonic matter is basically a quark matter, but with baryons near the 
quark Fermi surface \cite{McLerran:2007qj,McLerran:2008ua,Andronic:2009gj,Glozman:2007tv,Tsvelik:2021ccp,Hidaka:2008yy,Kojo:2009ha,Kojo:2010fe,Kojo:2011fh,Kojo:2011cn,Ferrer:2012zq}.
In the presence of the quark Fermi sea, the quark Pauli blocking effects force baryons to take large momenta, $P_B \sim \Nc \lqcd$ with $\lqcd \simeq 0.2-0.3$ GeV, 
because of $\Nc$ quarks collectively moving in the same direction. 
Remarkably, these baryons are relativistic at relatively low density, $n_B =1-4n_0 \sim \lqcd^3$,
much smaller than naive baryon gas estimates, $n_B \sim 100n_0 \sim P_B^3  \sim (\Nc \lqcd)^3$.
Including this feature leads to the peak in sound velocity around $n_B = 1-4n_0$
\cite{McLerran:2018hbz,Jeong:2019lhv,Duarte:2020xsp,Duarte:2020kvi,Zhao:2020dvu,Cao:2020byn,Margueron:2021dtx}.

Inspired by the work of McLerran and Reddy, 
one of the authors (T.K.) in this paper introduced a microscopic model of quark-hadron continuity \cite{Kojo:2021ugu}.
Although schematic, the model manifestly keeps track of the evolution of the occupation probability for quark states.
In dilute baryonic matter, quarks in isolated baryons are superpositions of momentum states from $0$ to $\sim \lqcd$, and each momentum state is occupied with a small probability.
Those quark states at low momenta are progressively occupied as baryon density increases, and at some density those states are fully occupied with the probability one.
We call it {\it quark saturation}.
It was found that, with a reasonable size scale for a quark wave function, the quark saturation happens at density considerably less than the density where baryon cores overlap.
This may be interpreted as a percolation of low momentum modes, referred as {\it soft deconfinement} in Ref.\cite{Fukushima:2020cmk}.
After saturating low momenta, the quark Fermi sea approaches a usual quark matter description.
In this description, the peak of sound velocity is triggered by the quark saturation effect,
while nuclear interactions tend to smooth out the peak structure. 
The description contains the above-mentioned quarkyonic matter model as a special case \cite{Kojo:2021ugu}.

While QC$_2$D differs from QCD with three colors in descriptions of baryons,   
QC$_2$D can be used to test the quark saturation effects;
the baryonic matter in dilute regime is the BEC of diquarks, but
such a picture breaks down as the quark occupation probability increases.
Also, lattice QC$_2$D studies found that diquark baryons have hard core repulsions \cite{Takahashi:2009ef}.
For these aspects, whether baryons are fermions or bosons may not be crucial,
and both cases should induce rapid stiffening associated with the quark saturation.
We delineate QC$_2$D from 
this point of view\footnote{We note that Ref.\cite{Hippert:2021gfs} also discussed the peak in sound velocity by focusing on the interplay between chiral and diquark condensates. Meanwhile, our scenario is not sensitive to the presence of chiral condensates.}
to find out concepts which are also useful for QCD.

Another important issue is the effects of interactions.
In the context of two solar mass NS, it is important to understand which interactions can stiffen quark matter EoS.
To get rough insights, it is useful to start with a simple parametrization of an energy density \cite{Kojo:2014rca,Baym:2017whm}
\beq
\varepsilon (n_B) = a n_B^{4/3} + b n_B^\alpha \,,
\label{eq:para_EOS}
\eeq
where the first term is from the relativistic kinetic energy for quarks, and the second is from interactions.
Taking a derivative to obtain $\mu_B$, using $P=\mu_B n_B - \varepsilon$, and eliminating the coefficient $a$,
one can write the pressure as
\beq
P = \frac{\, \varepsilon \,}{3} + b \bigg( \alpha - \frac{\, 4 \,}{3} \bigg) n_B^\alpha \,.
\label{eq:stiff_EoS}
\eeq
For $\alpha > 4/3$, repulsive ($b > 0$) forces stiffen EoS while attractive forces soften EoS. This trend is seen in low density nuclear physics.
Meanwhile, a less known possibility is $\alpha < 4/3$, for which attractive ($b<0$) forces stiffen EoS.
Such attractions with low powers of $n_B$ may happen around the Fermi surface at high density.
We argue that EoS becomes very stiff when the bulk part of the quark Fermi sea has effective repulsions but the Fermi surface enjoys the attractive correlations.
In QC$_2$D, the latter is naturally realized as diquark correlations.
Taking this schematic description as our baseline, we examine the effects of interactions using two models.
The first is a model based on coherent states of diquarks,
and the second is the Nambu-Jona-Lasinio (NJL) model with diquark pairing terms.

This paper is organized as follows.
In Sec.\ref{sec:quarks_in_BM} we discuss quarks in baryonic matter and the concept of the occupation probabilities.
In Sec.\ref{sec:ideal_gas} we compare ideal fermionic and bosonic baryon gases.
In Sec.\ref{sec:coherent_states} we discuss a model of coherent states for diquarks.
In Sec.\ref{sec:model} we study the NJL model.
Section \ref{sec:summary} is devoted to a summary.

\section{Quarks in baryonic matter}
\label{sec:quarks_in_BM}

In this section, we discuss quarks in a dense matter for a general number of colors.
We consider the occupation probability, $f_{\vq} = f_{q_s^c}$, of a quark state for a given set of color ($c$), flavor ($q$), spin ($s$) in baryonic matter.
We postulate the form  \cite{Kojo:2021ugu} ($\int_{\vp} \equiv \int \rmd^3 \vp/(2\pi)^3$),
\beq
f_{\vq} (\vk; n_B) =  \sum_I  \int_{\vK_B}\calB_I (\vK_B;n_B) Q^{I \vq }_{\rm in} (\vk, \vK_B) \,.
\label{eq:def_f_q}
\eeq
Here $\calB_I (\vK_B;n_B)$ is the number of baryons in a quantum state with a momentum $\vK_B$ and a baryon spin-flavor $(S,S_z, F)$, e.g., $\Delta^{++}_{S=S_z=3/2}$.
The function $Q^{I \vq}_{\rm in}$ is a single quark momentum distribution with a color-spin-flavor species $\vq = q^c_s = (u^R_\up, u^R_\down, u^G_\up,\cdots)$,
specifying the host baryon with a quantum number ``$I$''.
Since $f_{\vq}$ is the probability, it must obey the constraint
\beq
0 \le f_{\vq} (\vk ; n_B) \le 1 \,,
\eeq
which in turn constrains the form of $\calB$ \cite{Kojo:2021ugu}.
For the normalization of $Q^{I \vq }_{\rm in}$,
we set the spatial probability after momentum integration to be $1$. 
Then
\beq
\int_{\vk} Q_{\rm in}^{I \vq } (\vk, \vK_B)
 = \la I | N_{ q^c_s } | I \ra
 = \frac{1}{\, \Nc \,} \la I | N_{q_s} | I \ra \,,
\eeq
where $N_{q_s^c}$ is the number operator for a state $q_s^c$, and $N_{q_s} = \sum_c N_{q^c_s}$ 
which counts the number of states with a given spin-flavor ($q_s$).\footnote{
For example, at $\Nc=3$, a state $|  \Delta^{++}_{s_z=3/2} \ra = | u^R_\up, u^G_\up , u^B_\up \ra$ has the matrix element,
\beq
\la \Delta^{++}_{s_z=3/2} | N_{u_\up^R } |  \Delta^{++}_{s_z=3/2} \ra = 1
= \frac{\, \la \Delta^{++}_{s_z=3/2} | N_{u_\up } |  \Delta^{++}_{s_z=3/2} \ra \,}{3} \,,
\eeq
and $\int_{\vk} Q_{\rm in}^{\Delta^{++}_{s_z=3/2}  u^R_\up } (\vk, \vK_B) = 1$ for whatever $\vK_B$.
}
The total quark number is computed as
\beq
n_q  
&=& \sum_{q^c_s} \int_{\vk} f_{\vq} (\vk; n_B) 
\nonumber \\
&=&
 \sum_I \int_{\vK_B} \calB_I (\vK_B;n_B) \sum_{q^c_s} \la I | N_{q_s^c} | I \ra
 \nonumber \\
&=&
 \sum_I \int_{\vK_B} \calB_I (\vK_B;n_B) \sum_{q_s} \la I | N_{q_s} | I \ra
  \,,
  \label{eq:nq}
\eeq
where $\sum_{q_s} = \sum_{q_s^c}/\Nc$.
We note\footnote{For example, for a proton state $|p\ra \sim |uud\ra $ in QCD,
\beq
\la p | \sum_{q_s} N_{q_s} | p \ra = \la p |\big( N_{u_\up} + N_{u_\down} + N_{d_\up} + N_{d_\down}\big) | p \ra = 3 \,.
\eeq
} 
\beq
\sum_{q_s} \la I | N_{q_s} | I \ra = \Nc \,,
\eeq
with which Eq.(\ref{eq:nq}) leads to
\beq
\frac{\, n_q \,}{\, \Nc \,} = \sum_I \int_{\vK_B} \calB_I (\vK_B;n_B ) = \sum_I n_B^I = n_B \,,
\eeq
as it should.

As the baryon momentum is given by the sum of quark momenta,
$Q^{I \vq }_{\rm in}$ must satisfy the constraint\footnote{
As we integrate out $(\Nc-1)$-quarks in a baryon to get a single quark distribution $Q_{\rm in}$,  
the relation, $\vK_B = \vk_1 + \cdots + \vk_{\Nc}$, is casted onto a constraint on the average momenta,
$\vK_B=\Nc \la \vk \ra$. 
}
\beq
\vK_B = \Nc \int_{\vk} \vk Q^{I \vq }_{\rm in} (\vk,\vK_B) \,.
\eeq
where we assumed that $ Q^{I \vq }_{\rm in} $ is the same for all colors and simply multiplied a factor $\Nc$.
We further assume that the variance of quark momenta is characterized as
\beq
 \int_{\vk} \bigg( \vk - \frac{\, \vK_B \,}{\, \Nc \,} \bigg)^2 Q^{I \vq }_{\rm in} (\vk,\vK_B) \sim \lqcd^2 \,,
\eeq
for any $\vK_B$. 
The scale $\lqcd$ is about the inverse of a baryon size.
This constraint is satisfied by taking the following form,
\beq
Q_{\rm in}^{I \vq } (\vk,\vK_B) = Q^{I \vq}_{\rm in} \bigg( \vk - \frac{\, \vK_B \,}{\, \Nc \,}  \,, 0\bigg) \,.
\label{eq:Qin_Nc}
\eeq
If $|\vK_B| \ll \Nc \lqcd$, the term with $\vK_B$ can be treated as a small correction of $\sim 1/\Nc$.

\section{Ideal baryon gas}
\label{sec:ideal_gas}

In this section, we compare fermionic and bosonic baryon gases.
The purpose here is to examine to what extent the concepts of QC$_2$D can be applied for QCD.

\subsection{Fermionic baryon gas}
\label{sec:fermionic_baryon_gas}

We first consider baryons as fermions.
Let us discuss the dilute limit, $n_B \ll \lqcd^3$, of baryonic matter where baryons are supposed to form an ideal gas.
For baryons as fermions, we assumed that the number of spin-flavor states is $g_B = \sum_I 1$, and they are all degenerated. 
Then
\beq
\calB_I (\vk_B;n_B) = \theta( K_B^F - |\vK_B| ) \,,
\eeq
with 
\beq
 n_B = \sum_I \frac{\, (K_B^F)^3 \,}{\, 6\pi^2 \,}  \equiv g_B \frac{\, (K_B^F)^3 \,}{\, 6\pi^2 \,}  \,,
\eeq
where $K_B^F$ is the Fermi momentum of baryons defined through $n_B$.
To compute $f_{\vq }$, the discussions can be simplified by applying the $1/\Nc$ expansion to $Q_{\rm in}^{I \vq }$ [see Eq.(\ref{eq:Qin_Nc})]
\beq
f_{ \vq } (\vk; n_B) 
&=& \sum_I \int_{\vK_B} \calB_I (\vK_B;n_B) \bigg( Q_{\rm in}^{I \vq } (\vk, 0) + \cdots \bigg) 
\nonumber \\
&\simeq &~ 
 \frac{\, n_B \,}{\, g_B \,} \sum_I Q^{I \vq }_{\rm in} (\vk, 0) 
\,,
\label{eq:f_q_Nc0}
\eeq
where $\cdots$ are $1/\Nc$ corrections.
At leading order of $1/\Nc$, the function $f_{\vq}$ has the $\vk$ dependence as $Q^{I \vq }_{\rm in}$, but its magnitude grows linearly with $n_B$.
Recalling the constraint $f_{\vq } \le 1$, the ideal gas picture must be violated for some large $n_B$.
At leading order of $1/\Nc$, the saturation of the $\vk=0$ mode in a state $q_s$ takes place when 
\beq
n_B^{\rm id.sat} |_{\rm LO} = \bigg[ \frac{\, 1 \,}{\, g_B \,} \sum_I Q^{I \vq }_{\rm in} ({\bm 0}, 0) \bigg]^{-1} \,.
\eeq
From the normalization condition, $Q^{I \vq }_{\rm in} (0, 0) \sim \lqcd^{-3}$, $n_B^{\rm id.sat} \sim \lqcd^{3}$.
Beyond this critical density, the quark Fermi sea with the occupation probability $\simeq 1$ develops.

For simplicity, we estimate the energy density within the quasiparticle picture,
\beq
\varepsilon (n_B) =  \sum_{q^c_s} \int_{\vk} E_{ \vq } (\vk) f_{ \vq } ( \vk;n_B) \,,
\eeq
where $E_{\vq}$ is the single quark energy for states with $\vq=q^c_s$.
In a dilute limit, the leading order of $1/\Nc$ is
\beq
\varepsilon^{\rm LO}_{\rm dilute} (n_B) 
&=&  \frac{\, n_B \,}{\, g_B \,} \sum_I \sum_{q^c_s} \int_{\vk} E_{q^c_s} (\vk) Q^{I q^c_s}_{\rm in} (\vk, 0) 
\nonumber \\
&=&  \frac{\, n_B \,}{\, g_B \,} \sum_I M_B^I = n_B M_B
\,,
\eeq
where in the last step we used the degeneracy $M_B^I = M_B$ for all $I$.
For this energy density the pressure is vanishing
\beq
P^{\rm LO}_{\rm dilute} = n_B^2 \frac{\, \partial (\varepsilon_{\rm dilute}^{\rm LO} /n_B) \,}{\, \partial n_B \,}  = 0 \,.
\eeq
The pressure in a dilute limit is vanishing when a constant energy per particle is insensitive to changes in $n_B$.

This trend of the constant energy per particle must change when the quark saturation effects set in,
because the saturated levels require quarks to be added at higher momenta at larger $n_B$.
The energy per particle begins to grow just after the saturation, and the pressure increases rapidly.
Especially, if we naively extrapolate an ideal gas expression to the quark saturation point, 
the $n_B$ dependence of $\varepsilon/n_B$ changes discontinuously at the saturation.
This leads to an unphysical jump in pressure  \cite{Kojo:2021ugu}.
In more realistic treatments, at low density $\varepsilon/n_B$ increases more gradually mainly due to baryon-baryon interactions,
smearing the unphysical jump.

Throughout this section we have assumed that $ Q^{I \vq }_{\rm in} $ for quarks in a baryon does not change.
In principle, quark states in a baryon should be also affected when the low energy levels are partially filled.
This is also expected from meson exchanges between baryons that can be interpreted as quark hopping \cite{Fukushima:2020cmk}.
We see this sort of modification in the following sections.

\subsection{Bosonic baryon gas}
\label{sec:bosonic_baryon_gas}

In an ideal gas of elementary bosons, the ground state is the BEC made of bosons at zero momenta. 
For an ideal or noninteracting gas of composite bosons, 
we suppose that the BEC description is valid in a dilute regime.
Here, we consider only single bosonic baryon in two flavor theories,
namely a bosonic baryon made of a color-, flavor-, spin-singlet $ud$ diquarks.\footnote{
The quantum number can be expressed as
\beq
|B\ra = \epsilon_{cc'} \epsilon_{qq'} \epsilon_{ss'}\frac{\, |q^c_s, q'^{c'}_{s'} \ra \,}{\, \sqrt{2\Nc \Nf} \,} \,.
\eeq
The normalization factor is included in $Q_{\rm in}^{\vq}$.
}
In the BEC the baryon state distribution should be
\beq
\calB^{\rm ideal} (\vK_B;n_B) = n_B (2\pi)^3 \delta( \vK_B  ) \,,
\label{eq:ideal_def}
\eeq
which satisfies $n_B = \int_{\vK_B} \calB^{\rm ideal} (\vK_B;n_B) $.
The probability to find, e.g., the state $u^R_\up$ state in such a baryon is proportional to $1/2\Nf$ where the factor $2\Nf$ is from the spin and flavor $\Nf=2$.
Substituting Eq.(\ref{eq:ideal_def}) to Eq.(\ref{eq:def_f_q}), we can show 
\beq
f_{ \vq }^{\rm ideal} (\vk; n_B) \equiv n_B \, Q^{ \vq }_{\rm in} (\vk, 0) \,.
\label{eq:ideal_fq}
\eeq
which is similar to Eq.(\ref{eq:f_q_Nc0}).
The condition on a probability, $f_{ \vq } \le 1 $, is applied as before, so the saturation must take place at some density.
For an ideal gas
\beq
n_B^{\rm id.sat} = Q^{ \vq }_{\rm in} ({\bm 0}, 0)^{-1}  \sim  2\Nf \lqcd^3 \,.
\label{eq:id_sat}
\eeq
The EoS is soft ($P=0$) before reaching the saturation, and then gets stiffened.
This mechanism is the same as in a fermionic baryon gas.

\section{A model of coherent states}
\label{sec:coherent_states}

From this section we begin to take into account interactions.
For diquark baryons in QC$_2$D, relatively simple descriptions of $f_{\vq}$ from nuclear to quark matter regime are possible by referring to theories for 
the BEC-BCS crossover \cite{Leggett_book,schrieffer1999theory,Parish}.
In this section, we assume the presence of a hamiltonian which drives the formation of diquark bound states in vacuum. 
We do not write the hamiltonian explicitly, but assume that the resulting ground state can be described by the coherent states
with which the creation and annihilation operators of (composite) bosons have the nonzero expectation values, $\la d^\dag \ra, \la d \ra \neq 0$.
With these expectation values, the BEC in this section differs from the BEC of noninteracting bosons; 
the latter is the eigenstate of the number operator $d^\dag d$ leading to the zero expectation values of boson operators (e.g., $\la d \ra = 0$).
Hence, intrinsically the BEC in this section is made of interacting bosons.
For the stability of uniform Bose gas in dilute regime (no clustering), the boson-boson interactions should be repulsive here.
The pressure is nonzero from the dilute regime.

\subsection{Occupation probability}
\label{sec:occupation_probability}

In a dilute regime of interacting bosons, the BEC state can be described as a coherent state ($\calN$, $C$: constants to be fixed later)
\beq
| \Phi  \ra \equiv \calN \exp \big[ C \, b^\dag_0 \big] | 0\ra \,.
\eeq
We keep this form to the BCS regime.
Here $b_{\vK_B}^\dag$ is a creation operator of bosons with $\vK_B$, made of quark creation operators $\big( a^{q^c_s}_{\vk} \big)^\dag$ as ($\Nc = \Nf =2$)
\beq
b_{\vK_B}^\dag 
= \frac{\, \epsilon_{qq'} \epsilon_{cc'} \,}{\, \sqrt{\Nc \Nf} \,} \int_{\vk} \phi_{\vk} \, \big(a^{q^c_\up } _{\vk+\vK_B/2} \big)^\dag \big( a^{q'^{c'}_{\down} }_{-\vk+\vK_B/2} \big)^\dag \,.
\eeq
The function $\phi_{\vk}$ is a wave function for a relative momentum between two quarks,
normalized as $\int_{\vk} | \phi_{\vk} |^2 = 1$.
The form of $\phi_{\vk}$ may be density dependent in principle, but we use the same $\phi_{\vk}$ during the BEC-BCS crossover.
(In this section we neglect hole and antiparticle contributions for simplicity.)

Since the boson operator $b^\dag_0$ is made of fermions, we can rewrite the coherent state as ($\Phi_{\vk} \equiv C \phi_{\vk}$)
\beq
 | \Phi  \ra
&=& \calN \prod_{\vk}
 \exp \bigg[  \Phi_{\vk} \frac{\, \epsilon_{qq'} \epsilon_{cc'} \,}{\, 2 \,} 
 	 \big( a^{q^c_\up}_{\vk} \big)^\dag \big( a^{q'^{c'}_{\down} }_{-\vk} \big)^\dag \bigg] |0\ra 
	 \nonumber \\
&= & \calN \prod_{\vk} 
\, \sum_{n=0}^4 \frac{1}{\, n ! \,} \, \big( \Phi_{\vk} A_\vk^\dag \big)^n  |0\ra	 
\eeq
where 
\beq
A^\dag_{\vk} 
=  \frac{\, \epsilon_{qq'} \epsilon_{cc'} \,}{\, 2 \,}  \big(a^{q^c_\up } _{\vk} \big)^\dag \big( a^{q'^{c'}_{\down} }_{-\vk} \big)^\dag \,.
\eeq
In the product of $A^\dag_{\vk}$, the products of the same fermionic creation operators automatically drop off.
For given $(\vk \!\!\! \up; -\vk \!\!\! \down)$ levels, we have superposition of 
an empty state $|0; 0 \ra$, 
a diquark state (e.g., $|u^R_\up ; d^G_\down \ra$), 
a tetraquark state (e.g.,  $|u^R_\up  u^G_\up ; d^R_\down d^G_\down \ra$),
a hexaquark state, (e.g., $|u^R_\up u^G_\up d^R_\up ; u^G_\down d^R_\down  d^G_\down   \ra$)
and an octaquark state, $|u^R_\up u^G_\up d^R_\up d^G_\up ; u^R_\down  u^G_\down d^R_\down  d^G_\down   \ra$.
In the octaquark state all color-flavor-spin are filled for a given $\vk$;
states are saturated and no more states are available.
For our coherent states, all of these states have finite probabilities; the ground state changes from the BEC to BCS regime smoothly with changes in the weight factor $\Phi_{\vk}$.

In a dilute matter, $|\Phi_{\vk}| \ll 1$ for all levels and the diquark states are dominant.
In a dense matter, 
there are both dense and dilute levels in momentum space.
At low $\vk$, the octaquark states or filled quark Fermi sea are dominant with $|\Phi_{\vk}| \gg 1$.
But levels at sufficiently high $\vk$ have the small weight factors, $|\Phi_{\vk}| \ll 1$, leading to a smeared Fermi surface as in the BCS state.

Let us confirm the above-mentioned behaviors.
The normalization condition is
\beq
1
= \la \Phi | \Phi \ra 
= | \calN |^2 \prod_{\vk}  \sum_{n=0}^4 \frac{\,  \calA_n | \Phi_{\vk} |^{2n} \,}{\, (n !)^2 \,} \,,
\eeq
with the expectation values for each $\vk$,
\beq
 \calA_n \equiv \la 0 | A_\vk^n (A_\vk^\dag)^n | 0 \ra  \,.
\eeq
The list is
\beq
&&
\calA_0 = \calA_1 =1\,, ~~~ 
\calA_2 = \frac{\, 3 \,}{\, 2 \,} \,,~~~ 
\calA_3 = \calA_4 =  \frac{\, 9 \,}{\, 4 \,}  \,.
\eeq
Thus, the normalization constant $\calN$ is determined for a given set of $\Phi_{\vk}$ at various $\vk$.
The other constant $C$ is fixed by the condition ($n_q = 2n_B$)
\beq
\!\!\!\!\!\!
n_q 
= \sum_{q^c_s} \int_{\vk}   \la \Phi | n^{q^c_s}_{\vk} | \Phi \ra
= \sum_{q^c_s} \int_{\vk}  f_{\vq} (\vk;n_B) 
 \,,
 \label{eq:n_q_condition}
\eeq
where $n^{q^c_s}_{\vk} = \big( a^{q^c_s}_{\vk} \big)^\dag a^{q^c_s}_{\vk}$.
Explicit calculations lead to
\beq
 f_{ \vq } (\vk;n_B) 
&=& |\calN |^2 
\sum_{n=1}^4  \frac{\,  \calB_n | \Phi_{\vk} |^{2 n}  \,}{\, (n !)^2 \,}
\prod_{\vl \neq \vk}
 \sum_{m=0}^4 \frac{\,  \calA_m | \Phi_{\vl} |^{2m} \,}{\, (m !)^2 \,}
 \nonumber \\
&=&
 \sum_{n=1}^4  \frac{\,  \calB_n | \Phi_{\vk} |^{2 n}  \,}{\, (n !)^2 \,}
\bigg/ \sum_{m=0}^4 \frac{\,  \calA_m | \Phi_{\vk} |^{2m} \,}{\, (m !)^2 \,}
\,,
\eeq
with the expectation values for each $\vk$,
\beq
 \calB_n \equiv \la 0 | A_\vk^n \, n^{q^c_s}_{\vk} \, ( A_\vk^\dag)^n | 0 \ra  \,.
\eeq
The list is
\beq
&&
\calB_1 =  \frac{\, 1 \,}{\, 4 \,}\,, ~~~ 
\calB_2 =  \frac{\, 3 \,}{\, 4 \,} \,,~~~ 
\calB_3 =  \frac{\, 27 \,}{\, 16 \,} \,,~~~ 
\calB_4 =  \frac{\, 9 \,}{\, 4 \,}   \,.
\eeq
The constant $C$ is determined for a given set of the bound state wave functions $\phi_{\vk}$ at various $\vk$. 

In a dilute matter, $|\Phi_{\vk}|^2 \ll 1$ for all $\vk$, and we have 
\beq
n_q = 2n_B
\simeq \sum_{q^c_s} \int_{\vk} \calB_1 |\Phi_{\vk}|^2
= 2 |C|^2 
 \label{eq:C_nq}
 \,,
\eeq
where we have used $\int_{\vk} | \phi_{\vk} |^2 = 1$. Now we found $|C|^2 \simeq n_B$ with which 
\beq
 f_{ \vq } (\vk;n_B) 
\simeq \frac{\,  | \phi_{\vk} |^2 }{\, 4 \,} \, n_B
\,. ~~~~~~({\rm dilute~matter})
\label{eq:dilute_fq}
\eeq
The occupation probability is proportional to $n_B$.
The factor $1/4 = 1/2 \Nf$ reflects the probability to find specific spin-flavor quantum numbers (e.g., $u^R_\up$) from a spin- and flavor-singlet baryon. 

Meanwhile, in a dense matter, there are low momentum modes with $|\Phi_{\vk}|^2 \gg 1$ and high momentum modes with $|\Phi_{\vk}|^2 \ll 1$.
For low momentum modes, the expression of $f_{\vq}$ is dominated with the $|\Phi_{\vk}|^8$-terms. 
Here all color-flavor-spin quanta are saturated.
Including up to $1/|\Phi_{\vk}|^2$ correction, one finds (reminder: $\Phi_{\vk} = C \phi_{\vk}$)
\beq
f_{ \vq } (\vk;n_B) 
\simeq
 1 - \frac{\, 4 \,}{\, |\Phi_{\vk} |^2 \,}  \,.
\label{eq:dense_f}
\eeq
Thus $f_{\vq}$ approaches 1 from below, as it should.

Finally, we derive the expression for the amplitudes of diquark operators. 
It is given by 
\beq
 d_{\vk} 
 &\equiv&
  \la \Phi | A_{\vk} | \Phi \ra
= | \calN |^2 \prod_{\vk} \Phi_{\vk} \sum_{n=1}^4 \frac{\,  \calA_n | \Phi_{\vk} |^{2(n-1)} \,}{\, n! (n-1)!  \,} 
\nonumber \\
&=&
  \Phi_{\vk} \sum_{n=1}^4 \frac{\,  \calA_n | \Phi_{\vk} |^{2(n-1)} \,}{\, n! (n-1)!  \,} 
\bigg/ \sum_{m=0}^4 \frac{\,  \calA_m | \Phi_{\vk} |^{2m} \,}{\, (m !)^2 \,}
\,.
\eeq
Here $d_\vk$ is a dimensionless quantity. Diquark condensates are obtained after integrating $d_{\vk}$ over the phase space, $\int_{\vk}$.

Now, we try to gain the analytic insights. 
For dilute levels ($|\Phi_{\vk}|^2 \ll 1$),  
\beq
d_{\vk} ~\simeq~ \Phi_{\vk} \,.
\label{eq:di_dilute}
\eeq
In particular, if all levels are dilute, we may further use $|C|^2 \simeq n_B$ (Eq.(\ref{eq:C_nq})) to get $d _{\vk} \simeq \sqrt{n_B} \, \phi_{\vk}$.
On the other hand, for dense levels ($|\Phi_{\vk}|^2 \gg 1$),  
\beq
d_{\vk} ~\simeq ~ 4 \,\frac{\, \Phi_{\vk} \,}{\, |\Phi_{\vk}|^2 \,} ~ \simeq ~ \Phi_{\vk} \big[\, 1-f_{\vq} (\vk;n_B) \, \big]\,.
\label{eq:di_dense}
\eeq
In the last step we used Eq.(\ref{eq:dense_f}).
Thus, the diquark amplitudes are small for occupied levels.

\subsection{A model of gaussian wave functions}
\label{sec:gaussian_wf}

We now take a specific spatial wave function for composite bosons, 
and use them together with the formulas in Sec.\ref{sec:occupation_probability}.
We postulate the form
\beq
|\Phi_\vk |^2 = |C  \phi_\vk |^2 = |C|^2  \bigg(\frac{\, 2 \sqrt{ \pi } \,}{\, \Lambda \,} \bigg)^3 \, \rme^{ - \vk^2/\Lambda^2 } \,,
\label{eq:form_of_Phi}
\eeq
where $\Lambda$ characterizes the size of diquarks.
Taking the Fourier transform of $\phi_{\vk}$ and then squaring it, we obtain the probability of two quarks to have a relative distance $|{\bm r}|$,
\beq
|\phi_{\bm r}|^2 = {\rm const.} \times\rme^{-\Lambda^2 {\bm r}^2 } \,,
\eeq
from which one can calculate the average distance of two quarks in a baryon as
\beq
\la {\bm r}^2 \ra = 3 \Lambda^{-2} \frac{\, \int \rmd x \, x^2 \rme^{-  x^2} \,}{\,  \int \rmd x\, \rme^{- x^2}  \,} = \frac{3}{\, 2 \Lambda^2 \,}  \,.
\eeq
Thus, the average distance from the center is 
\beq
R \equiv \frac{\, \sqrt{ \la {\bm r}^2 \ra } \,}{2}
\simeq 0.612 \, {\rm fm} \times \bigg( \frac{\, 0.197\,{\rm GeV} \,}{\Lambda} \bigg) \,.
\eeq
This defines our baryon core radius in this paper.
We estimate the density where baryon cores overlap as
\beq
\!\!\!\!
n_B^{\rm overlap} 
\equiv \frac{\, 1 \,}{\, 4\pi R^3/3 \,}
\simeq 1.04 \, {\rm fm}^{-3}  \bigg( \frac{\Lambda}{\, 0.197\,{\rm GeV} \,} \bigg)^3 \,.
\eeq
For $\Lambda \simeq 0.2$ GeV, we find $n_B^{\rm overlap} \simeq 6.5 n_0$. 
It is also convenient to evaluate the density for the quark saturation at $\vk=0$.
Our measure is [see Eqs.(\ref{eq:id_sat}) or (\ref{eq:dilute_fq}) for $f_{\vq} (\vk=0;n_B)$]
\beq
n_B^{\rm id.sat} 
= 2\Nf  \bigg(\frac{\, \Lambda \,}{\, 2 \sqrt{ \pi } \,} \bigg)^3   
\simeq 0.09 \,{\rm fm}^{-3}  \bigg( \frac{\Lambda}{\, 0.197\,{\rm GeV} \,} \bigg)^3 \,.
\eeq
For $\Lambda \simeq 0.2$ GeV, we find $n_B^{\rm id.sat} \simeq 0.56 n_0$. 
This density is much smaller than that for baryons to  overlap.

\begin{figure}[tb]
\begin{center}	
\hspace{-0.8cm}
	\includegraphics[width=8.0cm]{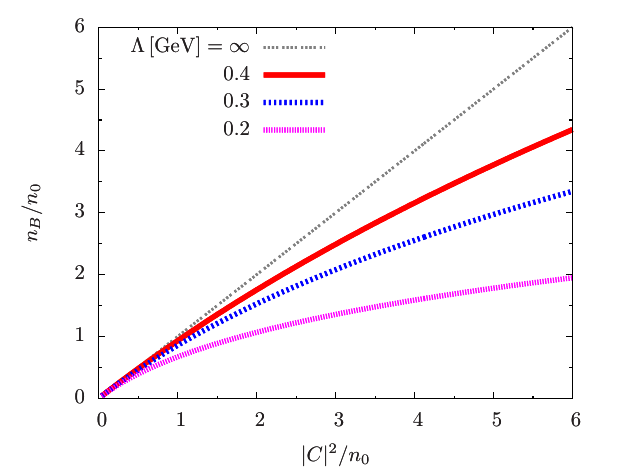}
\caption{ \footnotesize{$|C|^2$ vs $n_B$, each normalized by $n_0 = 0.16\,{\rm fm}^{-3}$. The value of $|C|^2$ corresponds to $n_B$ in the ideal Bose gas limit in which the composite nature is neglected or $\Lambda$ is taken to be infinity.
		} }
	\vspace{-0.5cm}
\label{fig:C2_vs_nb_n0}		
\end{center}
\end{figure}

Now, we first examine to what extent the ideal Bose gas expressions in Sec.\ref{sec:bosonic_baryon_gas} are valid.
We recall that, in the ideal gas limit, the parameter $|C|^2$ equals to $n_B$ [see Eq.(\ref{eq:C_nq})].
This equality is violated when the composite natures of baryons become important.
Shown in Fig.\ref{fig:C2_vs_nb_n0} is the $|C|^2$ vs $n_B$, each normalized by $n_0 = 0.16\,{\rm fm}^{-3}$.
The relations are naturally sensitive to $\Lambda$ or the baryon size.
The limit $\Lambda \rightarrow \infty$ corresponds to the point particle limit, and $|C|^2 = n_B$.
This simple relation does not hold for a smaller $\Lambda$;
 boson-boson interactions\footnote{
The effects of interactions are implicitly included by saying that the ground state is given by a coherent state with nonzero diquark ampliutdes.
}
as well as the composite nature of bosons make $n_B$ less than $|C|^2$. 
Remarkably, these effects are substantial at density much smaller than $n_B^{\rm overlap}$ and  $n_B^{\rm id.sat} $.
For instance, for $\Lambda=0.2$ GeV, the deviation from the ideal gas limit is significant around $n_B \simeq 0.5n_0$.
The coherent state is substantially different from the ideal Bose gas limit already at rather low density.

\begin{figure}[tb]
\begin{center}	
\hspace{-0.8cm}
	\includegraphics[width=8.0cm]{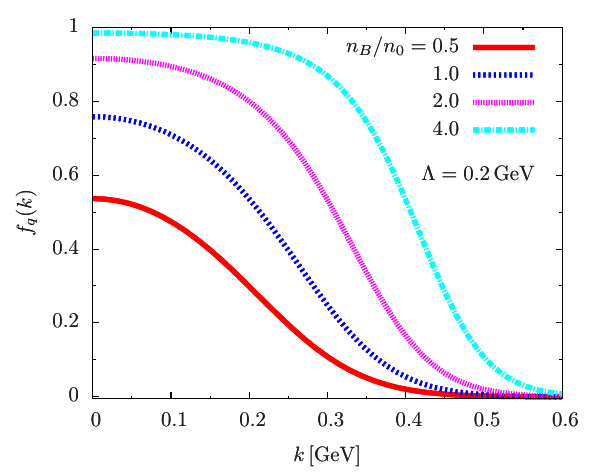}
\caption{ \footnotesize{ The occupation probability $f_q(k)$ as a function of momentum $k$. The cases $n_B/n_0 = 0.5, 1.0, 2.0$, and $4.0$ cases are shown.
		} }
	\vspace{-0.5cm}
\label{fig:fq_lam}		
\end{center}
\end{figure}

\begin{figure}[tb]
\begin{center}	
\hspace{-0.8cm}
	\includegraphics[width=8.5cm]{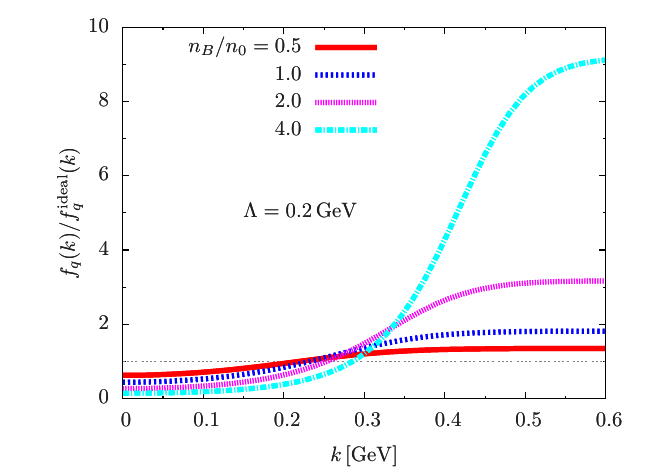}
\caption{ \footnotesize{  The $f_q(k)/f_q^{\rm ideal} (k)$ ratio for various $n_B$. The ratio measures how the quark Fermi sea differs from naive extrapolation of the ideal gas estimate.
		} } 
	\vspace{-0.5cm}
\label{fig:fq_phi2_lam}		
\end{center}
\end{figure}

Next, we examine the occupation probability of quark states, see Fig.\ref{fig:fq_lam} for the $\Lambda=0.2$ GeV case.
One can readily see that $\sim 50\%$ of the $k=0$ states are occupied at $n_B=0.5n_0$, and $\sim 100\%$ at $n_B =4n_0$.
At $n_B=0.5n_0$, one can see the Gaussian shape, but at higher density the shape is deformed.
The growth rate of the occupation probability becomes smaller for low momentum states,
and then states with higher momenta grow more substantially instead.
To see this trend more clearly, we examine the ratio
\beq
f_q(k)/f_q^{\rm ideal} (k)\,,
\eeq
where $f_q^{\rm ideal}$ is estimated by the expression of Eq.(\ref{eq:dilute_fq}). 
The result is shown in Fig.\ref{fig:fq_phi2_lam}.	
If we neglect the composite nature of baryons, the ratio remains $1$ for all momenta.
As density increases, the low momentum part of the ratio drops as the quark Pauli blocking effect tempers the growth in $f_q$,
and instead the higher momentum part in $f_q$ is increased more than in $f_q^{\rm ideal}$, leaving substantial enhancement in the ratio $f_q/f_q^{\rm ideal}$ at high momenta.
At $n_B\simeq 0.5n_0$, the occupation probability is considerably modified from the ideal gas limit.

\begin{figure}[tb]
\begin{center}	
\hspace{-0.8cm}
	\includegraphics[width=8.0cm]{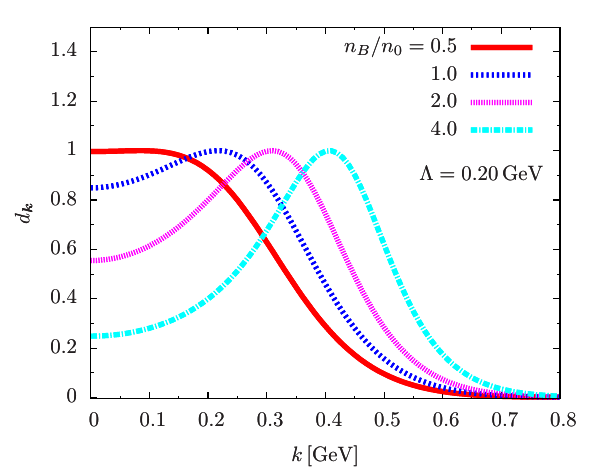}
\caption{ \footnotesize{ The (dimensionless) diquark amplitude $d_{\vk}$ as a function of momentum $k$. The cases $n_B/n_0 = 0.5, 1.0, 2.0$, and $4.0$ cases are shown.
		} }
	\vspace{-0.5cm}
\label{fig:di_lam}		
\end{center}
\end{figure}

Finally we examine a diquark amplitude $d_{\vk}$ which is dimensionless (Fig.\ref{fig:di_lam}).
From Eqs.(\ref{eq:di_dilute}) and (\ref{eq:di_dense}), its size is determined by the product of $\Phi_{\vk}$ and $1-f_\vq$.
For a large $k$, $\Phi_{\vk}$ becomes small but $1-f_{\vq} \simeq 1$.
For a small $k$, $\Phi_{\vk}$ is large but $1-f_{\vq}$ can be small at large $n_B$.
Hence, $d_{\vk}$ should take the maximum near the Fermi surface,
as clearly seen in Fig.\ref{fig:di_lam}.

\subsection{A model of single quark energy for EoS}
\label{sec:singel_E_EoS}

So far we have assumed the presence of a hamiltonian which favors the coherent states as ground states at given density,
but have not discussed the energies explicitly.
Here, we assume some form of average single particle energies and calculate the EoS within a quasiparticle picture.
We postulate the parametrization ($\beta > 0$)
\beq
E_{\rm CQM} (\vk) = \sqrt{ M_q^2 + \vk^2 } - C_A + C_S [ f_\vq (k) ]^\beta \,,
\label{eq:singe_q_E}
\eeq
where $M_q$ is the constituent quark mass, and
$C_A$ and $C_S$ are two-body semishort range correlations 
where $A$ and $S$ stand for color-singlet (antisymmetric) and triplet (symmetric) 
channels\footnote{
A more reasonable modeling should also distinguish spin-flavor channels. 
For a scalar diquark as discussed here, both color-electric and magnetic interactions act as attractions between two quarks, as in the case for pions in constituent quark models \cite{DeRujula:1975qlm}.
In this context, the values of $C_A$ should be interpreted as the sum of these two interactions.
}.
In a baryon or in a dilute matter, any two quarks are in the color-antisymmetric combination. Meanwhile, color-symmetric interactions occur only when two baryons are close and quarks from different baryons interact.
These pictures are reflected in our parametrization. In average, a quark has the energy reduction by $C_A$, while $C_S$ is suppressed by factors $ [ f_\vq (k) ]^\beta \ll 1$.
At a large density, quarks have more chance to interact with quarks in a color symmetric channel.
For occupied levels with $f_q(k)\sim 1$, the repulsive energy grows up.

With this single particle energy, the baryon mass is computed as the sum of average quark energy,
\beq
M_B 
&=& \Nc  \int_{\vk} E_{\rm CQM} (\vk) | \phi_{\vk}|^2   \nonumber \\
&=&  \Nc \bigg( \int_{\vk} | \phi_{\vk}|^2 \sqrt{ M_q^2 + \vk^2 }  - C_A \bigg) \,.
\eeq
Below we set $M_q=0.3$ GeV and $\Lambda=0.2$ GeV. 
For $C_A=0$, the baryon mass is $M_B\simeq 0.77$ GeV with $0.60$ GeV from the quark rest masses and 0.17 GeV from the quark kinetic energy.
We examine $C_A = 0.085, 0.185$, and $0.285$ GeV which give $M_B=0.6, 0.4$, and $0.2$ GeV, respectively.

For computations of EoS, we first set $C_S=0$.
We use the measures
\beq
\varepsilon^{\rm id. sat} \equiv M_B n_B^{\rm id. sat} \,,
\eeq
to estimate the energy density which the ideal gas picture would yield at the density of quark saturation.
For $C_A$'s to yield $M_B=0.6, 0.4$, and $0.2$ GeV, the corresponding  $\varepsilon^{\rm id. sat} $ are $0.054, 0.036$, and $0.018\,{\rm GeV fm^{-3}}$, respectively.
The corresponding pressure is zero because $\varepsilon/n_B$ is constant with respect to changes in $n_B$.

\begin{figure}[tb]
\begin{center}	
\hspace{-0.8cm}
	\includegraphics[width=9.0cm]{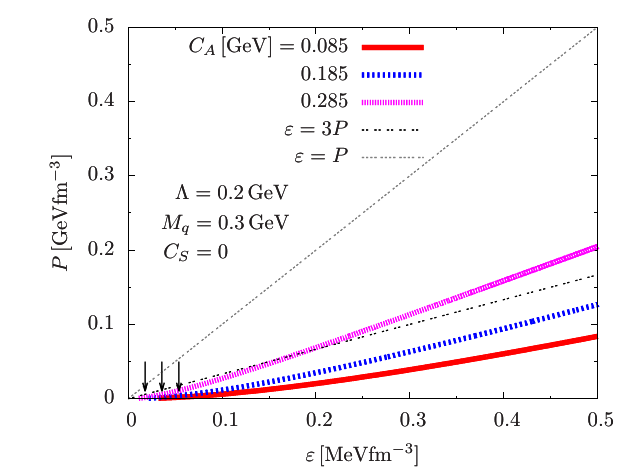}
\caption{ \footnotesize{  $\varepsilon$ vs $P$ for various $C_A$. We set $\Lambda=0.2$ GeV, $M_q=0.3$ GeV, and $C_S=0$. The arrows indicate the location of $\varepsilon^{\rm id. sat}=M_B n_B^{\rm id.sat}$. For $\Lambda=0.2$ GeV, the ideal gas picture should have the quark saturation at $n_B^{\rm id.sat} \simeq 0.56 n_0$.
		} } 
	\vspace{-0.5cm}
\label{fig:e-P_lam020_mq030}		
\end{center}
\end{figure}

\begin{figure}[tb]
\begin{center}	
\hspace{-0.8cm}
	\includegraphics[width=9.0cm]{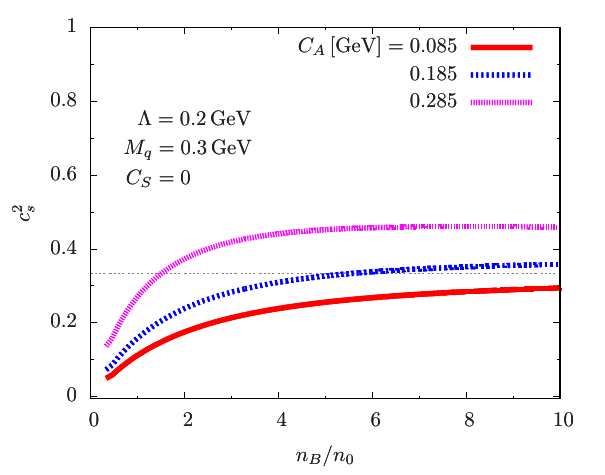}
\caption{ \footnotesize{ The square of sound velocity $c_s^2$ as a function of $n_B$. We set $\Lambda=0.2$ GeV, $M_q=0.3$ GeV, and $C_S=0$. The conformal value $c_s^2=1/3$ is also shown as a guide.
		} } 
	\vspace{-0.5cm}
\label{fig:nb_n0-cs2_lam020_mq030}		
\end{center}
\end{figure}

For these sets of parameters, the $\varepsilon$ vs $P$ is shown in Fig.\ref{fig:e-P_lam020_mq030}.	
First of all, all these curves suggest that the pressure at $n_B=n_B^{\rm id.sat}$ is not close to the ideal Bose gas limit, $P=0$.
Rather, the results of the coherent state smoothly approach the quark matter behaviors with the slope $\rmd P/\rmd \varepsilon = c_s^2$ close to $1/3$.
In particular, the speed of sound is already substantial before reaching $n_B^{\rm id. sat}$,
and no peaks are found (Fig.\ref{fig:nb_n0-cs2_lam020_mq030}).
We note EoS is stiffer for a larger $C_A$. 
This is because (with a given kinetic energy) a quark in attractive correlations can appear with the smaller energies.

Now, we turn on interactions in the color-symmetric channels, $C_S$, for $C_A=0.285$ GeV ($M_B=0.2$ GeV). 
We set $C_S=C_A$ for simplicity and vary the powers $\beta$ [see Eq.(\ref{eq:singe_q_E})].
The interaction part of a quark energy behaves as
\beq
-C_A + C_S  [ f_\vq (k) ]^\beta \sim -C_A \big( 1-  [ f_\vq (k) ]^\beta \big) \,.
\eeq
For a larger $\beta$, the contrast between the filled and empty levels is sharper.
Also, the effect of the repulsion is more suppressed ($ [ f_\vq (k) ]^\beta \ll 1$) at low density.

\begin{figure}[tb]
\begin{center}	
\hspace{-0.8cm}
	\includegraphics[width=9.0cm]{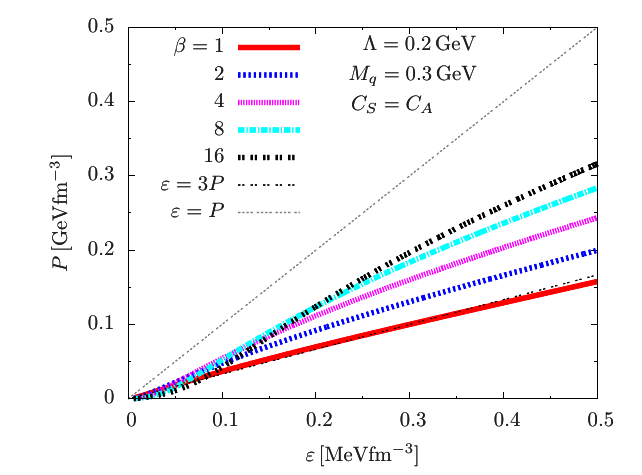}
\caption{ \footnotesize{ $\varepsilon$ vs $P$ for various $C_A$. We set $\Lambda=0.2$ GeV, $M_q=0.3$ GeV, and $C_S=0$. 
For $\Lambda=0.2$ GeV, the ideal gas picture should have the quark saturation at $\varepsilon \simeq 0.018\, {\rm GeV fm^{-3}}$ ($n_B^{\rm id.sat}= 0.56n_0$).
		} } 
	\vspace{-0.5cm}
\label{fig:e-P_lam020_mq030_beta_vary}		
\end{center}
\end{figure}

\begin{figure}[tb]
\begin{center}	
\hspace{-0.8cm}
	\includegraphics[width=9.0cm]{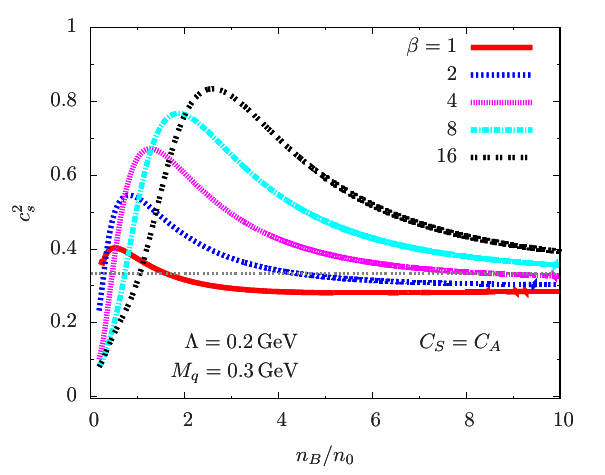}
\caption{ \footnotesize{ The square of sound velocity $c_s^2$ as a function of $n_B$. We set $\Lambda=0.2$ GeV, $M_q=0.3$ GeV, and $C_S=C_A$. The conformal value $c_s^2=1/3$ is also shown as a guide.
		} } 
	\vspace{-0.5cm}
\label{fig:nb_n0-cs2_lam020_mq030_beta_vary}		
\end{center}
\end{figure}

Shown in Fig.\ref{fig:e-P_lam020_mq030_beta_vary} is $\varepsilon$ vs $P$, and, in Fig.\ref{fig:nb_n0-cs2_lam020_mq030_beta_vary}, $n_B/n_0$ vs $c_s^2$.
The color-symmetric repulsion with $\beta\ge 1$ stiffens the EoS.
For a smaller $\beta$, the stiffening takes place at low density.
For a larger $\beta$, the repulsion sets in only when $f_q$ is very close to $1$.
The location of stiffening can be most clearly identified by peaks in $c_s^2$ (Fig.\ref{fig:nb_n0-cs2_lam020_mq030_beta_vary}).
Next, we examine why a larger $\beta$ leads to a stiffer EoS at high density.
For modes with $f_q (k) \simeq 1$, the powers $\beta$ no longer matter.
On the other hand, the correlations near the Fermi surface are sensitive to how much the repulsion is suppressed by the factor $ [ f_\vq (k) ]^\beta$.
For $C_A \gg C_S  [ f_\vq (k) ]^\beta$, the attractive correlations dominate over the repulsive one,
and the attractive correlations near the Fermi surface stiffens EoS, as advertised in Eq.(\ref{eq:stiff_EoS}).

\section{The NJL model}
\label{sec:model}

In this section, we discuss stiffening of matter in the two color and two flavor NJL model  \cite{Hatsuda:1994pi}.
The model can describe the chiral symmetry breaking/restoration, diquark condensations, the strucutral changes in hadrons, 
and so on \cite{Sun:2007fc,Ratti:2004ra,Brauner:2009gu,Andersen:2010vu,Andersen:2015sma,Imai:2012hr}.
Although the model does not include confining effects, baryons appear as composite particles whose masses in a dilute limit equal to the pion mass.
Thus, we expect that the stiffening mechanism in the previous sections has some validity.
We further elucidate various dynamical effects.

We consider light  doublet fields $ q =(u,d)^T$.
Our model Lagrangian is
\beq
\calL_{\rm NJL} 
= \bar{q} \big( \rmi \slash{\!\!\! \partial} - m + \mu \gamma_0 \big) q
 + \calL_4 \,,
\eeq
where $m=m_u=m_d$ is the current quark masses, 
and $\calL_4$ describes the chiral and diquark correlations among $(u,d)$ quarks,
\beq
\calL_4 
&=& 
G \big[\, (\overline{q} \tau_a q)^2 + (\bar{q} \rmi \gamma_5 \tau_a q)^2 \,\big]
\nonumber \\
&&
+
H \big[\, | \bar{q} \rmi \gamma_5 \tau_2 \sigma_2 q_C |^2
+ | \bar{q} \tau_2 \sigma_2 q_C |^2
\, \big] \,.
\eeq
This part of the Lagrangian is symmetric under $ U(\Nf)_L \otimes U(\Nf)_R \otimes SU(\Nc)$ transformations.
To check the symmetry for the last diquark terms, it is convenient to rewrite 
\beq
&&
 | \bar{q} \rmi \gamma_5 \tau_2 \sigma_2 q_C |^2
+ | \bar{q} \tau_2 \sigma_2 q_C |^2
 \nonumber \\
&& = 
2 |  \overline{q}_L \tau_2 \sigma_2 q_L^C |^2
 + 2 |  \overline{q}_R \tau_2 \sigma_2 q_R^C |^2 \,.
\eeq
In the last expression, each term is singlet in the $ U(\Nf)_L \otimes U(\Nf)_R \otimes SU(\Nc)$ group.
Note that the $(qq)_L (\bar{q} \bar{q})_R$ type cross terms are canceled so that the Lagrangian is $U(1)_A$ symmetric.
In this work we neglect the determinant interaction which breaks the $U(1)_A$ symmetry \cite{Hatsuda:1994pi,Brauner:2009gu}.
 
\subsection{Mean field thermodynamics}
\label{sec:MF}

In the mean field approximation,
\beq
&& \calL_4^{{\rm MF}}
= 4 G \sigma_f (\bar{q} q )_f
- 2 G  \sigma_f^2 
\nonumber \\
&&~~~~
 - H d ^\dag \left( \bar{q}_C \gamma_5 \tau_2 \sigma_2 q \right)
+ H d  \left( \bar{q} \gamma_5 \tau_2 \sigma_2 q_C \right)
- H |d |^2
\,,
\eeq
where the mean fields are given by
\beq
\sigma_f = \la ( \bar{q} q )_f \ra \,,
~~~~~~
d= \la \bar{q}_C \gamma_5 \tau_2 \sigma_{2} q \ra \,.
\eeq
The effective mass and diquark gap are given by
\beq
M_f = m_f - 4G \sigma_f \,,
~~~~~
\Delta = - 2 H d \,.
\eeq
Below we assume $M=M_u=M_d$.
Using the Nambu-Gor'kov bases, $\Psi = ( q, q_c )/\sqrt{2}$,
the mean-field Lagrangian can be written as
\beq
\calL_{\rm MF}
= \bar{\Psi} ( -\rmi S ) \Psi - 2 G  \sigma_f^2 - H |d |^2 \,,
\eeq
where $S$ is the Nambu-Gor'kov propagator,
\begin{align}
-\rmi S^{-1} 
= \left(\begin{array}{cc}
~~ \rmi \Slash{\partial} - \hat{M} + \mu \gamma_0 ~&~
  \Delta \gamma^5 \tau_2 \sigma_2 ~\\
~ -  \Delta^* \gamma^5 \tau_2 \sigma_2 ~&~
 \rmi \Slash{\partial} - \hat{M} - \mu \gamma_0 ~\,
\end{array}
\right) 
\,,
\end{align}
The mean field thermodynamic potential is ($T$: temperature, $\mu$: quark chemical potential)
\beq
\Omega_{\rm MF}
&=& 
 \Nc \Nf
 \sum_{\xi = {\rm p, a} }
 \int_\vk
\left[\, - \epsilon^\xi_{\vk}  - 2T \ln \left(\,1+\rme^{- \epsilon^\xi_{\vk} /T } \,\right)\,
\right] 
\nonumber \\
&&~~+
 2 G  \sigma_f^2 + H |d |^2 \,.
\eeq
with the quasiparticle energies
\begin{eqnarray}
\epsilon^{\xi}_{\bm k} &=& \sqrt{(E_{\bm k}- \eta_\xi \mu)^2+|\Delta|^2} \,, 
\end{eqnarray}
where we introduced $\eta_{\rm p} = +1$ and $\eta_{\rm a} = -1$ and $E_{\bm k} = \sqrt{ {\bm k}^2 + M^2}$.
As measures of occupation probability, it is common to use the functions
\begin{eqnarray}
|u_{\xi}({\bm k}) |^2 &=& \frac{1}{\, 2 \,} \left(1+ \frac{\, E_{\bm k}- \eta_\xi \mu \,}{\, \epsilon^{\xi}_{\bm k} \,} \right)\ ,  \nonumber\\
|v_{\xi}({\bm k}) |^2 &=& \frac{1}{\, 2 \,} \left(1 - \frac{\, E_{\bm k}- \eta_\xi \mu \,}{\, \epsilon^{\xi}_{\bm k} \,} \right)\ , 
\end{eqnarray}
which satisfy 
\begin{eqnarray}
\!\!\!\!
|u_{\xi}({\bm k}) |^2+|v_{\xi}({\bm k}) |^2 = 1\,,~~
 u_{\xi}({\bm k})  v_{\xi}({\bm k})  = \frac{\Delta}{2\epsilon_{\xi}({\bm k}) } \,.
\end{eqnarray}
The baryon number density $n_B^{\rm MF} = n_q^{\rm MF}/\Nc$ with $n_q^{\rm MF} = - \partial \Omega_{\rm MF}/\partial \mu$ is
\beq
n_q^{\rm MF} = 2 \Nf \int_{\vk} \big(\, f^{\rm MF}_{\rm p}(\vk) - f^{\rm MF}_{\rm a} (\vk) \, \big) \,,
\eeq
where $f_{\rm p,a}^{\rm MF}$ is the quark and antiquark occupation probabilities,
\beq
f_\xi^{\rm MF} (\vk) 
&\equiv& 
 |v_\xi (\vk) |^2 	\big( 1-2n_\vk^\xi \big) + n_\vk^\xi 
 \,,
 \label{eq:f_q^MF}
\eeq
with the Ferm-Dirac distribution function
\beq
n_\vk^\xi = \frac{1}{\, 1+\rme^{ \epsilon^\xi_{\vk} /T} \,} \,.
\eeq
In the zero temperature limit, $n_\vk^\xi \rightarrow 0$.

\subsection{Numerical analyses}
\label{sec:numerical}

We now numerically examine the mean field EoS.
We use the following set of parameters:
\beq
\Lambda = 1.0\,{\rm GeV}\,,~~~ G \Lambda^2= 2.8\,,~~~ m_q= 5\,{\rm MeV}\,,
\eeq
which yield the vacuum pion mass $m_\pi \simeq 0.17$ GeV and the vacuum constituent quark mass $M\simeq 0.31$ GeV.
We restrict ourselves to the $T=0$ case.
We first present the bulk results for EoS and then analyze the microscopic aspects.


\begin{figure}[tb]
\begin{center}	
\vspace{-0.5cm}
	\includegraphics[width=8.8cm]{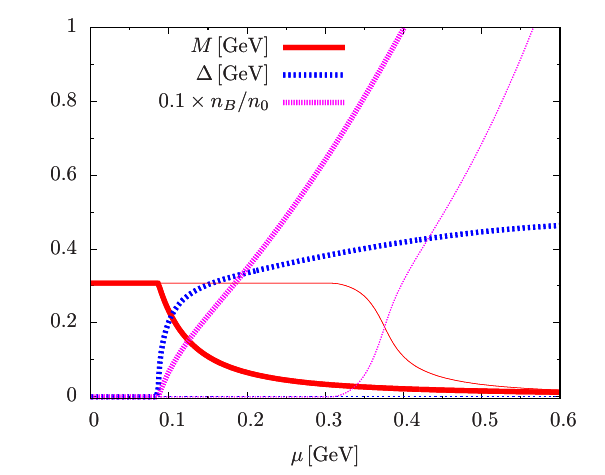}
\caption{ \footnotesize{The Dirac mass ($M$), diquark gap ($\Delta$), and normalized baryon number density $n_B/n_0$ with $n_0=0.16\,{\rm fm}^{-3}$.
	The thin lines are the results for $H=0$ with which $\Delta =0$. 
		} }
	\vspace{-0.5cm}
\label{fig:mass}		
\end{center}
\end{figure}

\begin{figure}[tb]
\begin{center}	
\hspace{-0.8cm}
	\includegraphics[width=9.2cm]{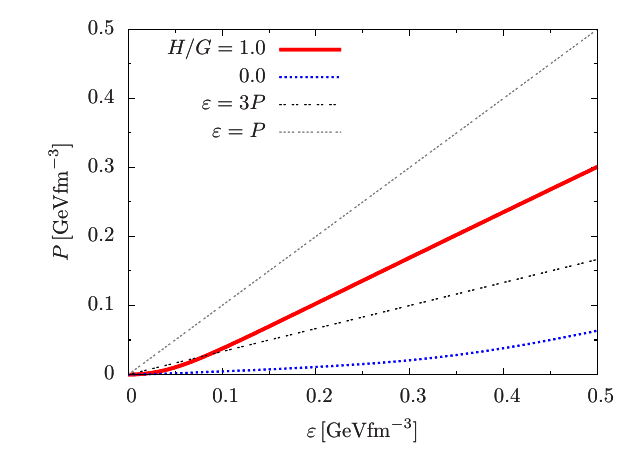}
\caption{ \footnotesize{The $\varepsilon$ vs $P$ for $H/G=1.0$ and $0.0$. The results for $\varepsilon=3P$ and $P$ are also shown as guide lines. 
		} }
	\vspace{-0.5cm}
\label{fig:e-p}		
\end{center}
\end{figure}

\begin{figure}[tb]
\begin{center}	
\vspace{-0.1cm}
\hspace{-0.45cm}
	\includegraphics[width=8.5cm]{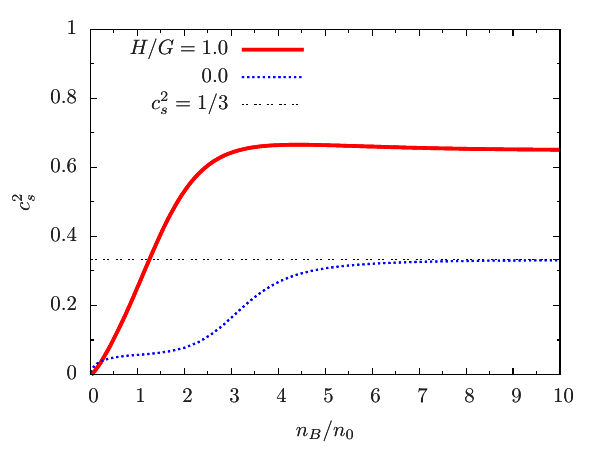}
\caption{ \footnotesize{The $n_B/n_0$ vs $c_s^2$ for $H/G=1.0$ and $0.0$. 
		} }
	\vspace{-0.5cm}
\label{fig:nb-cs2}		
\end{center}
\end{figure}

Shown in Fig.\ref{fig:mass} are the Dirac mass ($M$), diquark gap ($\Delta$), and normalized baryon number density ($n_B/n_0$) where $n_0=0.16\,{\rm fm}^{-3}$ 
for $H=G$ (bold lines) and $H=0$ (thin lines).
The onset of baryon density is sensitive to the parameter $H$.
For $H=G$, the baryon density is brought by the diquark baryon with the mass $m_\pi$, and
the corresponding quark chemical potential is $\mu_q = \mu_B/2 = m_\pi/2$.
For $H=0$, there are no composite baryons, and the baryon density becomes nonzero at $\mu_q = M$.

Shown in Figs.\ref{fig:e-p}	is  $\varepsilon$ vs $P$, which measures the stiffness of EoS.
As guidelines, we also show EoS with $\varepsilon = 3P$ and $\varepsilon = P$ for which $c_s^2 = 1/3$ and $1$, respectively.
For $H=G$, the pressure is substantial at relatively small $\varepsilon$.
This is in sharp contrast with the $H=0$ case where the pressure develops very slowly with increasing $\varepsilon$.
With diquark attractions, quarks can appear with small energies at a given pressure, as discussed in Sec.\ref{sec:singel_E_EoS}.

The rapid stiffening can be characterized by the behavior of the sound velocity, $n_B/n_0$ vs $c_s^2$, as shown in Fig.\ref{fig:nb-cs2}.
The sound velocity for $H=G$ exceeds the conformal limit at relatively low density, $n_B/n_0 \simeq 1.2$, reaches the maximum at $n_B/n_0 \sim 4$,
and gradually decreases.
In contrast, for $H=0$ the sound velocity very slowly grows with density and approach the conformal value $c_s^2 =1/3$ from below.

%



\begin{figure}[tb]
\begin{center}	
\vspace{-0.1cm}
\hspace{-0.45cm}
	\includegraphics[width=8.5cm]{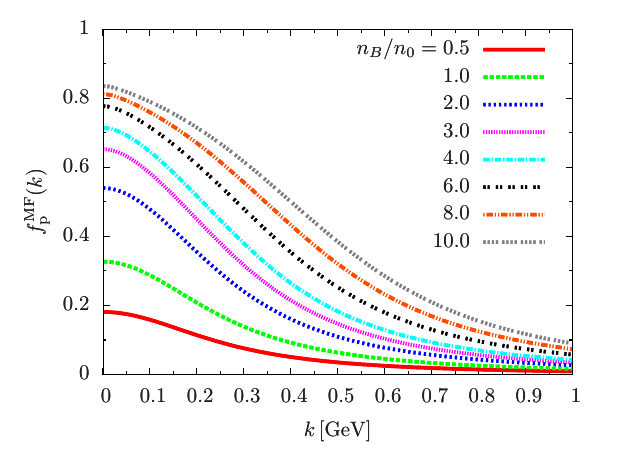}
\caption{ \footnotesize{The occupation probability of particle states, $f^{\rm MF}_{\rm p} (k)$, for various $n_B/n_0$. 
		} }
	\vspace{-0.5cm}
\label{fig:nb-fp}		
\end{center}
\end{figure}

\begin{figure}[tb]
\begin{center}	
\vspace{-0.5cm}
\hspace{-0.45cm}
	\includegraphics[width=8.5cm]{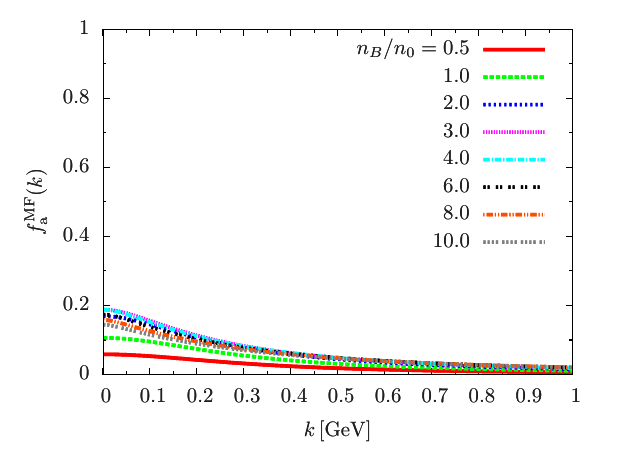}
\caption{ \footnotesize{The occupation probability of antiparticle states, $f^{\rm MF}_{\rm a} (k)$, for various $n_B/n_0$. 
		} }
	\vspace{-0.5cm}
\label{fig:nb-fa}		
\end{center}
\end{figure}

As we have seen in our model of coherent states, 
the states with high momenta are occupied substantially even before the saturation of low momentum states is completed.
The behavior of $f_{\rm p}^{\rm MF} (k)$ is shown in Fig.\ref{fig:nb-fp}.
For example, from $n_B/n_0=0.5$ to $1.0$, the occupation probability of the $k=0$ state already does not scale as $\sim n_B$.
With increasing $n_B$, the states at high momenta are occupied more rapidly than in the ideal gas case,
and accordingly $\varepsilon/n_B$ and hence $P$ grow.

Finally we take a look at the occupation probability for antiparticle states.
In the NJL model, the gap $\Delta$ is common for particle and antiparticle states.
As a consequence the impacts of $\Delta$ on antiparticle states are not small.
The occupation probability initially increases as $\Delta$ does, 
but at $n_B/n_0 \simeq 3$ the antiparticle suppression by large $\mu$ eventually dominates over the effects from $\Delta$.

%

\section{Summary}
\label{sec:summary}

We have discussed stiffening of matter in QC$_2$D within the coherent state model and NJL model.
In both models diquark baryons appear as composite particles.
A dense matter of composite objects observes quarks as their constituents, 
and is subject to the quark saturation effects that stiffen EoS.
Here, we summarize our findings, their implications for QCD, and future works.

(i) The quark saturation effects affect diquark baryons at density significantly lower than the overlapping density of baryon cores.
For diquarks with the radii $\sim 0.6$ fm, the saturation effects are substantial already at $n_B \sim 0.5-1 n_0$.
In three color QCD, this density range is regarded as the territory of conventional nuclear physics where the matter properties depend on intricate balance among various nuclear forces.
It is interesting to ask whether using quark descriptions simplifies some part of the nuclear matter descriptions at $n_B \sim n_0$.

\begin{figure}[tb]
\begin{center}	
\vspace{-0.4cm}
\hspace{-0.45cm}
	\includegraphics[width=8.5cm]{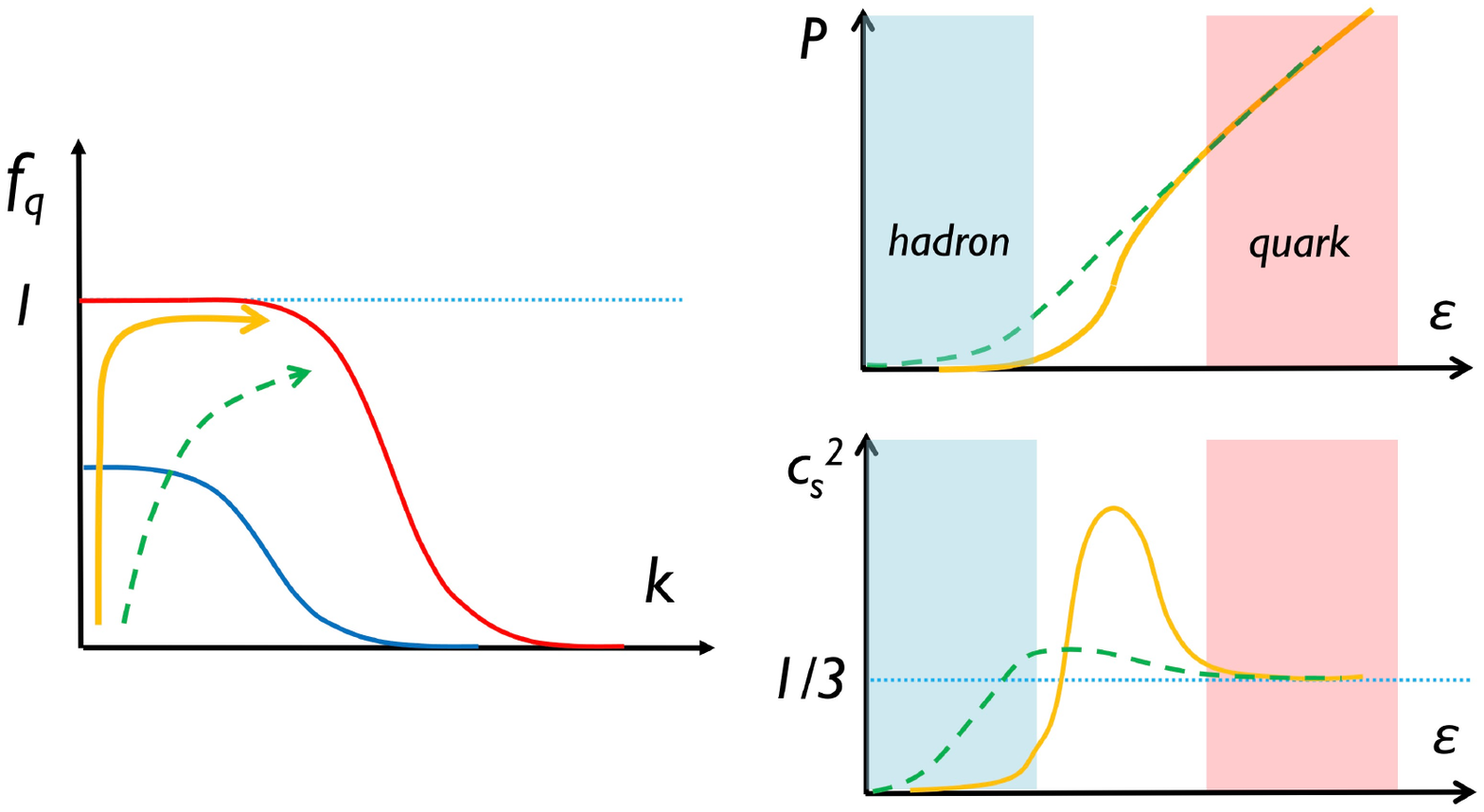}
	\vspace{-0.5cm}
\caption{ \footnotesize{The relations among $f_\vq (k)$ and EoS. When $f_{\vq}$ evolves as in an ideal baryon gas picture, then the quark saturation radically stiffens EoS and makes a peak in the sound velocity $c_s^2 = \rmd P/\rmd \varepsilon$. If the baryon gas picture breaks down at lower density, stiffening mildly occurs from low density and the peak structure is modest or absent.
		} }
\label{fig:fq_evolution_cs2}		
\end{center}
\end{figure}

(ii) In the coherent state model, a baryonic matter quickly gets stiffened and approaches the quark matter regime at low density.
With this small disparity between baryonic and quark matter regimes, the quark saturation effects alone lead to only a rather mild peak in the sound velocity.
The peak becomes more prominent when baryonic matter stays soft before the quark saturation effects set in (Fig.\ref{fig:fq_evolution_cs2}).
In QCD, nuclear matter at $n_B\simeq n_0$ is soft, and in this quantitative aspect the three color case somewhat differs from baryonic matter in QC$_2$D.

(iii) In a dilute regime, attractive correlations inside a baryon, which reduces the baryon mass, stiffens EoS. 
This is simply because the energy density tends to be smaller for a smaller baryon mass.
In terms of the parameterized EoS in Eq.(\ref{eq:para_EOS}) of Sec.\ref{sec:Introduction}, here we are considering the case with $b<0$ and $\alpha=1$.
The same feature was also found in the three color case \cite{Kojo:2021ugu}.

(iv) In a dense regime, attractive correlations near the Fermi surface stiffens EoS.
This has been discussed in models for QCD \cite{Kojo:2014rca,Kojo:2015fua,Fukushima:2015bda,Baym:2017whm,Baym:2019iky}, and we also reassure its impact in QC$_2$D.

(v) The best combination to stiffen EoS is to have attractive correlations near the Fermi surface but repulsive correlations in the bulk part of the Fermi sea, 
as demonstrated in this work and also in the three color studies \cite{Kojo:2021ugu}. 
This surface-bulk disparity on the sign of interactions may be explained by perturbative gluon exchanges; 
they lead to the color electric interactions, which overall reduces the energy of color-singlet states,
and magnetic interactions,
which reduces the energy for color- and spin-singlet states \cite{DeRujula:1975qlm}.
In a dilute regime, a matter is dominated by states in attractive correlations, 
but in a denser regime repulsive channels are also unavoidable.
This channel dependence leads to the above-mentioned disparity.

We believe that many aspects discussed in this paper, except the quantative details of baryonic matter, are common for QC$_2$D and QCD. 
This work is largely based on effective models and focus on the transition regime.
Meanwhile, it is also interesting to study the high density regime where weak coupling methods may be 
applicable \cite{Freedman:1977gz,Freedman:1976ub,Annala:2019puf,Gorda:2021kme,Gorda:2021znl,Fraga:2015xha,Kurkela:2014vha,Kurkela:2009gj,Fujimoto:2020tjc,Song:2019qoh}.
There QC$_2$D and QCD may be similar and it may be possible to derive a constraint on QCD EoS from the high density side.

Finally, we close this paper by mentioning a significant advantage to use a single framework from baryonic to quark matter regimes.
While implicit, our descriptions did not allow the liberty to add a bag constant by hand.
In a single model, there are no questions about the {\it normalizations of EoS} for baryonic and quark matter,
or no subtleties about the subtraction of the vacuum or Dirac sea contributions \cite{Kojo:2018usw}.
In QC$_2$D, a single model description was possible as baryons are just two particles.
The same program should be carried out for QCD by handling three particle correlations from nuclear to quark matter regime.
Attempts toward this goal are in progress.

\section*{Acknowledgment}

T. K. thanks Dr. Hatsuda for his questions on the sound velocity in the BEC-BCS crossover,
Dr. Yasutake for discussions during a NS workshop at RIKEN,
and Drs. Tachibana and Hidaka for discussions during a workshop on the thermal field theory. 
T. K. is supported by NSFC grant No. 11875144.

\bibliography{ref}

\end{document}